\documentclass[
]{ceurart}

\usepackage{etoolbox}
\makeatletter
\patchcmd{\@copyrightLine}{\textcopyright}{Copyright \textcopyright}{}{}
\makeatother
\usepackage{listings}
\begin{document}

\copyrightyear{2026}
\copyrightclause{for this paper by its authors. Use permitted under Creative Commons License Attribution 4.0 International (CC BY 4.0).}

\conference{Proceedings of the Joint Ontology Workshops (JOWO) - Episode XII: The Tropical Spring of Ontology, co-located with the 16th
International Conference on Formal Ontology in Information Systems (FOIS 2026), September 21-22, 2026, Vitória (ES), Brazil}

\title{BZKO: An Ontology for the Card Index of German Post-War Compensation Records}


\author[1]{Dilek Yargan}[%
orcid=0000-0001-9618-6740,
email=Dilek.Yargan@fiz-Karlsruhe.de,
]
\cormark[1]
\address[1]{FIZ Karlsruhe -- Leibniz Institute for Information Infrastructure,  
Hermann-von-Helmholtz-Platz 1, 76344 Eggenstein-Leopoldshafen, Germany}

\author[1]{Jörg Waitelonis}[%
orcid=0000-0001-7192-7143,
email=Joerg.Waitelonis@fiz-Karlsruhe.de,
]

\author[1,2]{Mahsa Vafaie}[%
orcid=0000-0002-7706-8340,
email=Mahsa.Vafaie@fiz-Karlsruhe.de,
]

\author[1,2]{Harald Sack}[%
orcid=0000-0001-7069-9804,
email=Harald.Sack@fiz-Karlsruhe.de,
]
\address[2]{Karlsruhe Institute of Technology,  
Institute of Applied Informatics and Formal Description Methods,  
Kaiserstr. 89, 76133 Karlsruhe, Germany}

\cortext[1]{Corresponding author.}

\begin{abstract}
The Central Federal Card Index (\textit{Bundeszentralkartei}) of Germany is a key archival resource documenting compensation claims submitted by victims of National Socialist persecution and their relatives, within the German \textit{Wiedergutmachung} process. To enable semantically enriched representation, integration, and reuse of this historically significant collection, we present the BZK Ontology (BZKO). 
We propose a two-layer ontology for historical archival data that separates ontologically grounded domain semantics from interoperability-oriented extension constructs. The approach combines BFO-based realism with archival standards (RiC-O, PROV-O, PiCo), enabling provenance-preserving semantic integration, while maintaining logical rigor, modularity, and reuse across digital humanities infrastructures.
The proposed approach establishes a reusable semantic foundation for the integration of \textit{Wiedergutmachung} archival materials into digital humanities infrastructures and lays the groundwork for future knowledge graph generation, ontology validation, and the incorporation of additional historical entities and uncertain temporal and spatial information.

The ontology is available on \url{https://github.com/ISE-FIZKarlsruhe/bzko}. 
\end{abstract}


\maketitle

\section{Introduction}

The historical process of \textit{"Wiedergutmachung nationalsozialistischen Unrechts" (Wiedergutmachung for National Socialist Injustice)}, left behind voluminous amount of compensation records that constitute one of the most historically significant archival collections related to post-war Germany. \textit{Wiedergutmachung}, literally "making good again", refers to the reparation and restitution process initiated following the fall of the National Socialist regime in Germany, which between 1933 and 1945 systematically persecuted groups of people on the grounds of ethnicity, religion, political belief, and sexual orientation among others \cite{BFM}. The archival collections related to \textit{Wiedergutmachung} comprise over 100 km of archival items, such as index cards, applications for compensation, restitution, and hardship provisions, as well as administrative files. Although not all the materials are available, either because they are not in digitized form or because there are restrictions under archival law for a subset of the cards, to make the documentary legacy of Wiedergutmachung accessible for searching for persons or groups, organizations, and scientific research, the German Federal Ministry of Finance has initiated the project of \textit{Themenportal Wiedergutmachung}, a central digital portal for the collection of \textit{Wiedergutmachung} files from various state and federal archives in Germany and beyond \cite{ArchivPortal, BMFArchProject}. To this end, the description and digitalization of archival material, using various information extraction and semantic structuring technologies, have been required to facilitate accessibility and semantic search. 

The Central Federal Card Index (\textit{Bundeszentralkartei}, abbreviated BZK) is a card index containing over 1.9 million cards, known as \textit{BZK index cards}, or \textit{BZK cards} for short. A BZK card is an administrative record created by a clerk at a compensation office after a person applied for compensation, mostly under the Federal Compensation Act (\textit{Bundesentschädigungsgesetz}, BEG for short) \cite{BezirkDuesseldorf}. It contains key personal and procedural details, such as name, birthplace, residential address, date of death, the name of the responsible compensation authority, and a reference number, together with a range of additional personal and administrative details, such as marital status, nationality, information on heirs or surviving dependents, and the claimed type of compensation. The BZK cards have then served as an internal work instrument for compensation authorities, registering compensation applications, facilitating central tracking and cross-referencing of cases across different authorities. In the future, as the BZK itself has now become archival material, these cards will serve as the hub of the other archival materials in the collections, and making them available for the public is of the highest importance for \textit{Themenportal Wiedergutmachung}.

This paper introduces the BZKO (BZK Ontology), designed to semantically represent relevant data from BZK index cards and to serve as the foundation for a knowledge graph that supports semantic search on \textit{Themenportal Wiedergutmachung}. The paper is organized as follows. Section 2 provides an overview of the data contained on the cards to be modeled. Section 3 presents prior work related to modeling and ontology design in the context of the Wiedergutmachung project, and Section 4 details the methodology and tools used to model the ontology. Section 5 suggests the classes, relations, and shortcuts for modeling the data on BZK cards. Section 6 discusses the findings, and Section 7 concludes by sketching plans for future work.

In this paper, class names are written in italics along with their corresponding namespaces, e.g., \textit{bfo:entity}. Relations, together with their namespace, are written in bold, e.g., \textbf{rdfs:subClassOf}. Lastly, the logical connectors are in small all-caps, e.g., \textsc{some}. Resource-specific naming conventions are preserved, such as the use of PascalCase or camelCase e.g., \textit{picom:PersonObservation} or \textbf{rico:hasOrHadIdentifier}. 

The ontology is publicly available at GitHub on https://github.com/ISE-FIZKarlsruhe/bzko. 

\section{Information Recorded on the BZK Index Cards}

The BZK cards served originally as an internal tool that were created when the case entered the compensation system to identify which compensation claim was filed by whom, on behalf of whom, and which compensation authority handled the case. Thus, the information recorded on the cards was primarily used to identify individuals and cases. Now, the index cards themselves, as archival material, will provide the first basis for semantic search on \textit{Themenportal Wiedergutmachung}.

Vafaie et al.\cite{vafaie2025end} introduced BZKOpen \cite{BZKopen}, a dataset designed to serve as a ground truth, i.e., the correct reference data for measuring the performance of information extraction algorithms. Its distinct fields were manually annotated by the domain experts from the Federal Archives of Germany. The dataset consists of 516 digitized BZK cards, curated to respect privacy constraints while preserving a diverse range of layout styles. This manually annotated, high-quality sample has been used in this study to build and evaluate the BZK Ontology. 

\begin{figure}
  \centering
  \includegraphics[width=\linewidth-90pt]{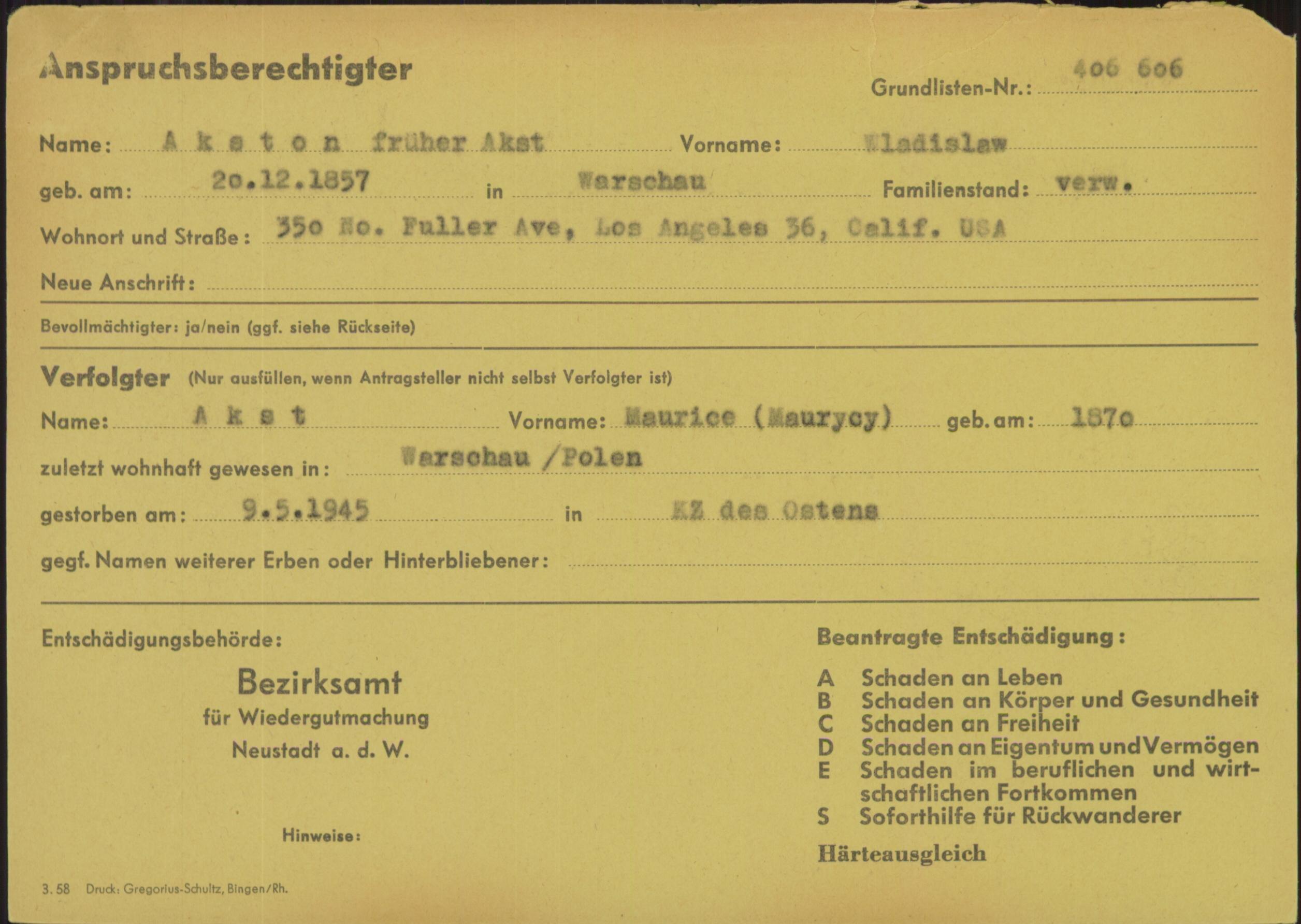}
  \caption{A sample of a BZK card from BZKOpen \cite{BZKopen}. Compensation office: Bezirksamt für Wiedergutmachung Neustadt a. d. W.; BZK number: 406 606; BZK card layout: RLP-Hauptphase; Applicant’s first name: Wladislaw; Applicant’s last name: Akston; Applicant’s birth name: Akst; Applicant’s date of birth: 1857-12-20; Applicant’s birthplace: Warschau; Applicant’s residential address: 390 No. Fuller Ave, Los Angeles 36, Calif. USA; Persecutee’s first name: Maurice; Persecutee’s last name: Akst; Persecutee’s alternative first name: Maurycy; Persecutee’s date of birth: 1870; Persecutee’s last residential address: Warschau/Polen; Persecutee’s date of death: 1945-05-09; Persecutee’s deathplace: KZ des Ostens. Source: LAV NRW R, BR 3015.}
    \label{fig:sampleBZKcard}
\end{figure}

The information recorded on the cards was primarily used to identify individuals, including applicants for compensation and persons persecuted by the Nazi regime, and organizations (more details in Section 4). As such, the cards may include the name of the compensation office, the BZK number (a reference number used for tracking compensation claims), applicant’s and/or persecutee’s first name, last name, birth name, and other alternative names, date of birth, birthplace, persecutee’s date of death, deathplace, and last residential address (see Figure~\ref{fig:sampleBZKcard}). Some cards contain more information about the applicant, the persecutee, and the heirs, such as their marital status or residential address, while others lack structured sections for registering information about the applicant or the persecutee. 

\section{Related Work}

Within the context of the \textit{Wiedergutmachung} project and in direct relation with the \textit{Wiedergutmachung} collection, there is some prior research with a focus on modeling and ontology design. Modeling archival collections on the Semantic Web requires ontologies that can jointly address archival structure, provenance, and the historical actors and events embedded in the records. 

In \cite{vafaie2021modelling}, Vafaie et al. address the challenge of representing heterogeneous archival documents and justify the reuse of the Records in Contexts Ontology (RiC-O)\footnote{https://www.ica.org/standards/RiC/RiC-O\_1-1.html} to model the \textit{Wiedergutmachung} archival documents. They argue that RiC-O's approach of modeling archival hierarchy levels as named individuals offers a more generic strategy that broadens its applicability across archives of different historical periods and geographical contexts. They also point out that RiC-O's inclusion of finer-grained entities within the hierarchy improves findability and allows for a more precise representation of archival resources.

In CourtDocs \cite{CourtDoc_paper}, the authors propose a data model to represent the \textit{Wiedergutmachung} process, starting with the application for compensation. This data model addresses the challenges of representing  relationships between documents, the agents involved in their creation, and the procedural events they describe, in a semantically coherent and interoperable way. Heterogeneous document types created in legal proceedings, as well as different administrative and judiciary steps are represented in this data model, drawing on established ontologies, namely RiC-O, the Provenance Ontology (PROV-O)\footnote{https://www.w3.org/TR/prov-o}, and the CIDOC Conceptual Reference Model (CRM)\footnote{https://cidoc-crm.org}.

\section{Two-Layer Ontology Design}

To represent the content of the cards, the BZK Ontology (BZKO) is currently built as an application ontology and initially developed in a bottom-up manner. Although a BZK card can contain a wide range of information, such as an applicant’s name and date of birth, their marital status, compensation office that issued the claim, and the names of the representative, as an initial step, we have collected all features from the BZKOpen database \cite{BZKopen} that constitute the main entities of the ontology. In order to support interoperability, standardization, and ontology reuse, we built upon two established ontologies. The Basic Formal Ontology (BFO), a top-level ontology standardized in ISO/IEC 21838-1 \cite{Arp2015BFO, ISOBFO, BFOGithub}, ensures philosophical and formal consistency at scale. It thereby provides clear distinctions between material objects and information artifacts, roles and dispositions, processes, and description-logics-based formality, which are often intentionally blurred in widely used digital humanities tools, such as CIDOC CRM, for practical reasons. Using BFO, thus, minimizes ambiguity in modeling decisions and promotes consistent modeling. NFDIcore, a BFO-based mid-level ontology developed for representing metadata related to the German National Research Data Infrastructure (NFDI), was selected to ensure interoperability within the NFDI ecosystem \cite{NFDI, NFDICoreGithub}. NFDIcore reuses existing ontologies, including the Information Artifact Ontology (IAO), the Relations Ontology (RO), and the Ontology for Biomedical Investigations (OBI). 

Moreover, we have extended the ontology by aligning it with and importing from the following semantic resources. The Provenance Ontology (PROV-O), a W3C standard, is mainly used to represent and exchange provenance information consistently. Thus, its use enables data traceability across cards and supports accountability. The Records in Contexts Ontology (RiC-O), as mentioned above, is used to represent and manage archival records. Its classes and relations enable the semantically enriched representation of BZK cards and their digitized form in archival terminology. Although RiC-O and NFDIcore have classes to represent persons identified in the BZK cards, we imported the classes from Persons in Context (PiCo)\footnote{https://personsincontext.org} to extend semantic representations of persons, as it is specifically designed to model historical person data and genealogical reconstructions. Lastly, Schema.org\footnote{https://schema.org} data properties are imported, as this semantic resource provides broad interoperability through a widely recognized vocabulary. It thereby enhances interoperability and facilitates data integration. 

BZKO adopts a two-layer architecture. The \textit{knowledge layer} contains all the necessary and sufficient classes and relations to model the information recorded on a BZK card. The classes in this layer are primitive (in description logics terminology) and rigid (in OntoClean terminology) classes. In other words, they are the building blocks for representing the content of the BZK cards. Their essential characteristics persist throughout their existence, and they compose other BZKO-related classes. For instance, the "BZK card" class is introduced as a primitive class, since it represents an entity that cannot be defined using existing BZKO classes. It is also considered rigid; its instances cannot cease to be a BZK card without ceasing to exist.

The \textit{extension layer} contains imports from the above-mentioned semantic resources, as well as the defined and non-rigid classes and relations of BZKO. Accordingly, the release artifact of the knowledge layer, \texttt{bzk.owl}\footnote{https://github.com/ISE-FIZKarlsruhe/bzko/blob/main/bzk.owl}, includes imports from BFO and NFDIcore, which subsume the primitive classes of BZKO. The artifact of the extension layer, \texttt{bzk-extension.owl}\footnote{https://github.com/ISE-FIZKarlsruhe/bzko/blob/main/bzk-extension.owl}, includes all classes and relations of BZKO and shortcuts and mappings based on the mentioned semantic resources (See Section~\ref{nonrigid} and Section~\ref{extension layer}).

The rationale behind the architecture is twofold. The two-layer structure improves reusability by enabling the knowledge layer to be used independently across different applications. For instance, the knowledge layer of BZKO can be imported on its own and extended with RiC-O, without requiring the extension layer or its PiCo-based classes. This approach also improves maintainability by ensuring that changes to the knowledge layer are propagated in a controlled manner, preventing circular modifications.
\begin{figure}
  \centering
  \includegraphics[width=\linewidth-145pt]{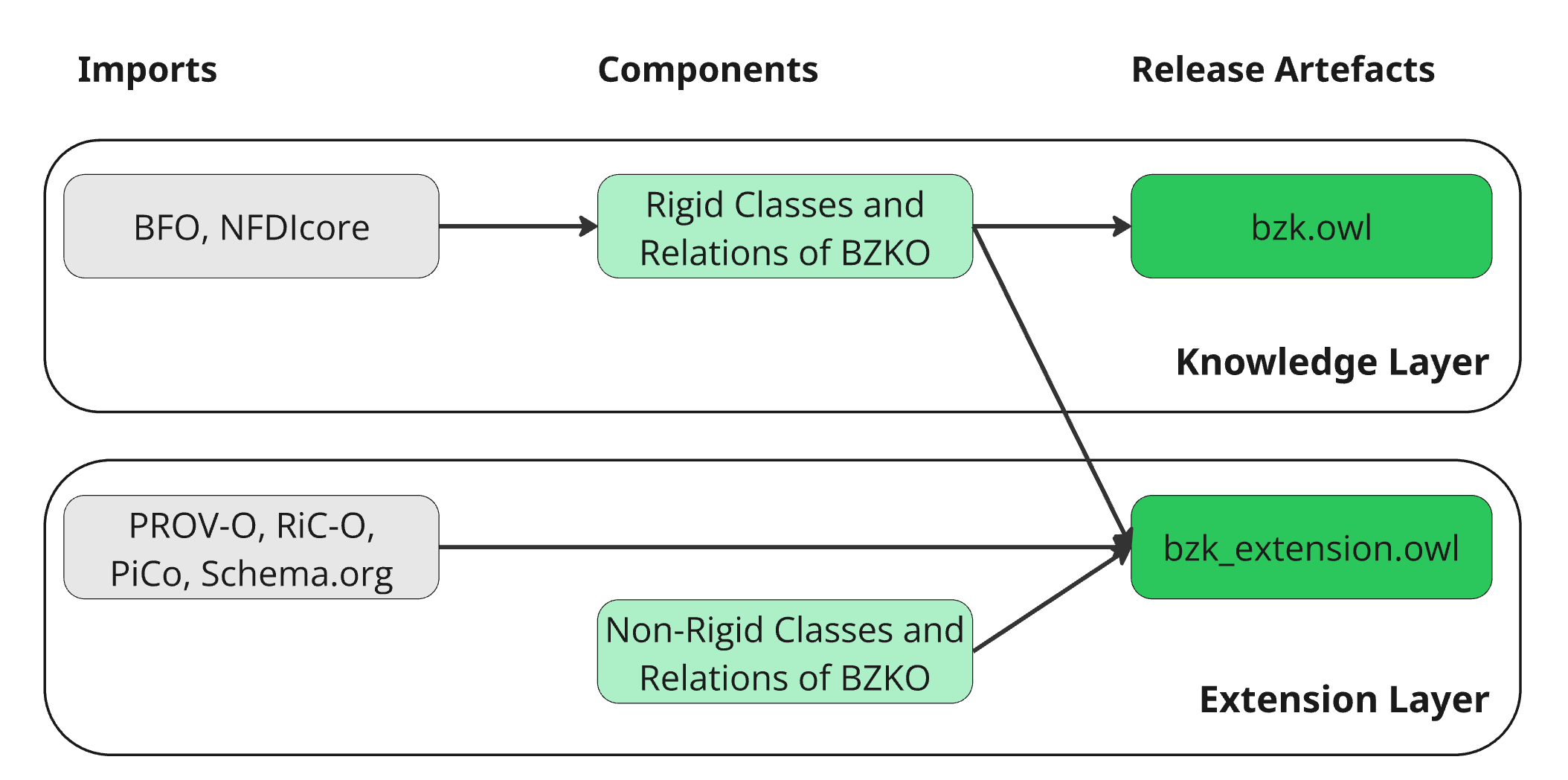}
  \caption{Two-layer design of BZKO, with each layer corresponding to a distinct release artifact. Arrows denote dependency and inclusion relations: each target incorporates or depends on its source.}
    \label{fig:BZKOlayers}
\end{figure}

Figure~\ref{fig:BZKOlayers} illustrates the two-layer architecture of BZKO and its main structural components. Each layer is characterized by three elements: imported semantic resources, components, and a corresponding release artifact. The components of the layers were determined based on the ontological statuses of features in BZKOpen \cite{BZKopen}. Each feature was examined to determine the most specific BFO and NFDIcore class, or classes, where compositional modeling was required, under which it could be appropriately subsumed. Then, each feature was represented either by a single ontology class or by a composition of classes. For instance, the feature "BZK number" was represented by \textit{bzko:bzk number}, since its being an identifier that persists over time in this domain; whereas, "applicant" was modeled through the composition of \textit{bzko:applicant role}, \textit{nfdi:person}, and \textbf{ro:has role}, since roles are dependent on the bearer and are realized in processes: a person can have multiple roles (applicant and persecutee roles) at once or hold a role (applicant) at different times. 

Each class and relation was then analyzed to determine whether it could be defined in terms of other classes and relations in BZKO. The primitive classes and relations constitute the components of the knowledge layer, while the others, together with their formal equivalents, either constitute the components of the extension layer or are represented through shortcut relations. Finally, to provide mappings from domain-related semantic resources, relevant external classes were incorporated into the imports of the extension layer.

We are compiling three collections of competency questions in collaboration with experts from the Federal Archives of Germany and the State Archives of Baden-Württemberg. The first collection includes organizational questions that define the scope of the ontology and support the validation of terminology choices. The second collection includes validation questions that assess the ontological correctness with respect to its logical structure. The third collection includes knowledge acquisition questions derived from use cases to evaluate whether the ontology meets project requirements and constraints. The answers of the first collection have been reflected in the ontology design, namely, the coverage of the ontology was approved by the domain experts.  On the other hand, the last two are ongoing work. The evaluation of these collections requires data; therefore, they were excluded from the current evaluation and deferred to future work. 

For ontology development, we follow the Open Biological and Biomedical Ontology (OBO) Foundry Principles \cite{OBO}, a collection of best practices aimed at supporting orthogonality and interoperability across ontologies \cite{OBOFoundry2021}. The ontology has been implemented in OWL (Web Ontology Language), using Protégé 5.6.7. 

The Ontology Development Kit (ODK) has been used to manage ontology development, dependency imports, and quality control in a reproducible manner \cite{ODK_paper, ODKGitHub}. External classes and object properties were imported from reference ontologies using ODK's import pipeline, which extracts only the required terms together with their relevant axioms while preserving references to the source ontologies with their particular versions. This approach minimizes redundancy and facilitates interoperability. Different ontology variants were managed through the ODK build workflow, enabling the automatic generation of release artifacts and import modules for distinct use cases. Ontology quality was assessed using the validation tools integrated into ODK, including automated reasoning and consistency checks, as well as ontology quality-control tests to identify missing annotations, unsatisfiable classes, and structural inconsistencies prior to release.

\section{Results}
In the following subsections, the components of each BZKO layer and their relations to the imported classes and relations are discussed. 

\subsection{The Knowledge Layer of BZKO}\label{knowledgelayer}
The knowledge layer includes classes and relations from BFO and NFDIcore, as well as classes and relations imported from ontologies such as IAO, OBI and RO, which are included in BFO and NFDIcore. It also includes rigid classes of BZKO, which are necessary for defining domain knowledge in the BFO framework.

A BZK index card, a physical archival object, records information about agents, names, dates, places, and identifiers. Its corresponding representation, \textit{bzko:bzk card}, is necessary to establish the provenance of the data. A digitized version of a physical BZK card has been used to extract information, so its corresponding representation is necessary not only to represent the domain correctly within the BFO framework, but also because the file names of the digitized cards have been used in date-extraction procedures, as they are the dates of birth of either applicants or persecutees. Hence, \textit{bzko:digitized bzk card} is also introduced in the ontology. 

During the indexing process, most BZK cards were assigned an identifier, called the BZK number, which is a Federal Central Register number referring to a compensation claim. This number may be used to track the corresponding compensation files in the archives holding them today. Consequently, the introduction of \textit{bzko:bzk number} is necessary.

The content of a card includes information about agents, namely, about persons (applicants and persecutees) and an organization (a compensation office). In this context, an applicant is a person who was recorded on a BZK card and applied for compensation for a persecuted person or for themselves; a persecutee, on the other hand, is a person who was subjected to persecution by the Nazi regime, including but not limited to discrimination, dispossession, forced displacement, or violence, on political, racial, religious, or other grounds. A compensation office is a government agency that reviews claims and provides compensation to persecuted people. In order to model these agents within the BFO framework, we need to define their roles, as being an applicant or a compensation office is not per se. These roles depend on a process, which realizes them. For this reason, we introduce \textit{bzko:compensation card indexing}, a subclass of \textit{bfo:process}, along with \textit{bzko:applicant role}, \textit{bzko:persecutee role}, and \textit{bzko:compensation office role}. 

The class \textit{bzko:compensation card indexing}, however, is associated with roles that are not actually realized through the indexing process: the applicant role is realized through participation in the compensation application process, while the persecuted-person role is realized through relevant persecution-related processes. In our model, therefore, we intentionally adopt a coarser level of granularity. The purpose of the BZK card is to identify persons as applicants or persecuted persons and organizations as compensation offices. The card-indexing process records the attribution of these roles to the corresponding agents rather than realizing the roles themselves.

A collection of information on the cards contains the names of the agents. On the one hand, representing the names of compensation offices is straightforward, i.e., the name of a compensation office instantiates \textit{bzko:compensation office name}; on the other hand, representing the names of persons requires a more fine-grained modeling. The first and family names of persons are represented by \textit{iao:given name} and \textit{iao:family name}, respectively. Additionally, the family name a woman had before marriage is represented by \textit{bzko:birth name}. In the context of BZKO, \textit{bzko:alternative given name} and \textit{bzko:alternative family name} represent given names and family names that have been altered or rendered in German or Latin forms, respectively. However, none of these classes in the ontology is domain-specific. 

A distinct collection of information on the cards concerns dates and places. Both applicants and persecutees have dates of birth and birthplaces, but some of the persecuted people have dates of death and death places. To represent dates and places, \textit{bzko:birth process boundary} and \textit{bzko:death process boundary} are introduced \cite{Arp2015BFO}. How to represent birthplaces, deathplaces, and dates of birth and death based on these process boundaries will be defined in Section 4.2.

Addresses constitute another collection of information on the cards. The class \textit{iao:postal address} is used to represent the addresses of persons during BZK card creation, and \textit{bzko:last residential} address is introduced to represent the last known voluntarily addresses of persons voluntarily during persecution. 

There is one class left, which is neither the content of a card nor necessary for the ontological representation of content within the BFO framework: \textit{bzko:bzk card layout}. A layout type refers to the classification of BZK index cards by experts at the Federal State Archive, based on the spatial distribution of information across the card layout. For instance, the layout of the BZK card shown in Figure~\ref{fig:sampleBZKcard}, labelled \textit{ RLP-Hauptphase}, provides distinct, structured sections for both the applicant and persecutee, but the layout type \textit{NRW-Innenministerium} lacks such sections. This entity is required in the modeling, as criteria for determining the layout types have been used in the information extraction processes.

\subsection{Non-Rigid Classes and Relations of BZKO}\label{nonrigid}
A class is \textit{non-rigid} (in the OntoClean sense \cite{OntoClean}) if its instances can cease to instantiate it without ceasing to exist. For instance, a person can cease to \emph{be} an applicant while continuing to \emph{exist} as a person. This distinction important in BZKO: roles, places, and dates are all non-rigid, as they depend on a specific context or process.

The first group of non-rigid BZKO classes concerns agents. The people whose names appear on the BZK cards hold applicant and/or persecutee roles within the Wiedergutmachung context. For this reason, they are represented by the classes \textit{bzko:wgm applicant} and \textit{bzko:persecutee of the nazi regime}. The following are examples of defined classes in BZKO.
\begin{itemize}
  \item[] \textit{bzko:persecutee of the nazi regime} \textbf{rdfs:subClassOf} \textit{nfdi:person} \textsc{and} (\textbf{ro:has role} \textsc{some}  \textit{bzko:persecutee role})
  \item[] \textit{bzko:compensation office} \textbf{rdfs:subClassOf} \textit{obi:organization} \textsc{and}  (\textbf{ro:has role} \textsc{some} \textit{bzko:compensation office role})
\end{itemize}

\begin{figure}[ht]
  \centering
  \includegraphics[width=\linewidth-16pt]{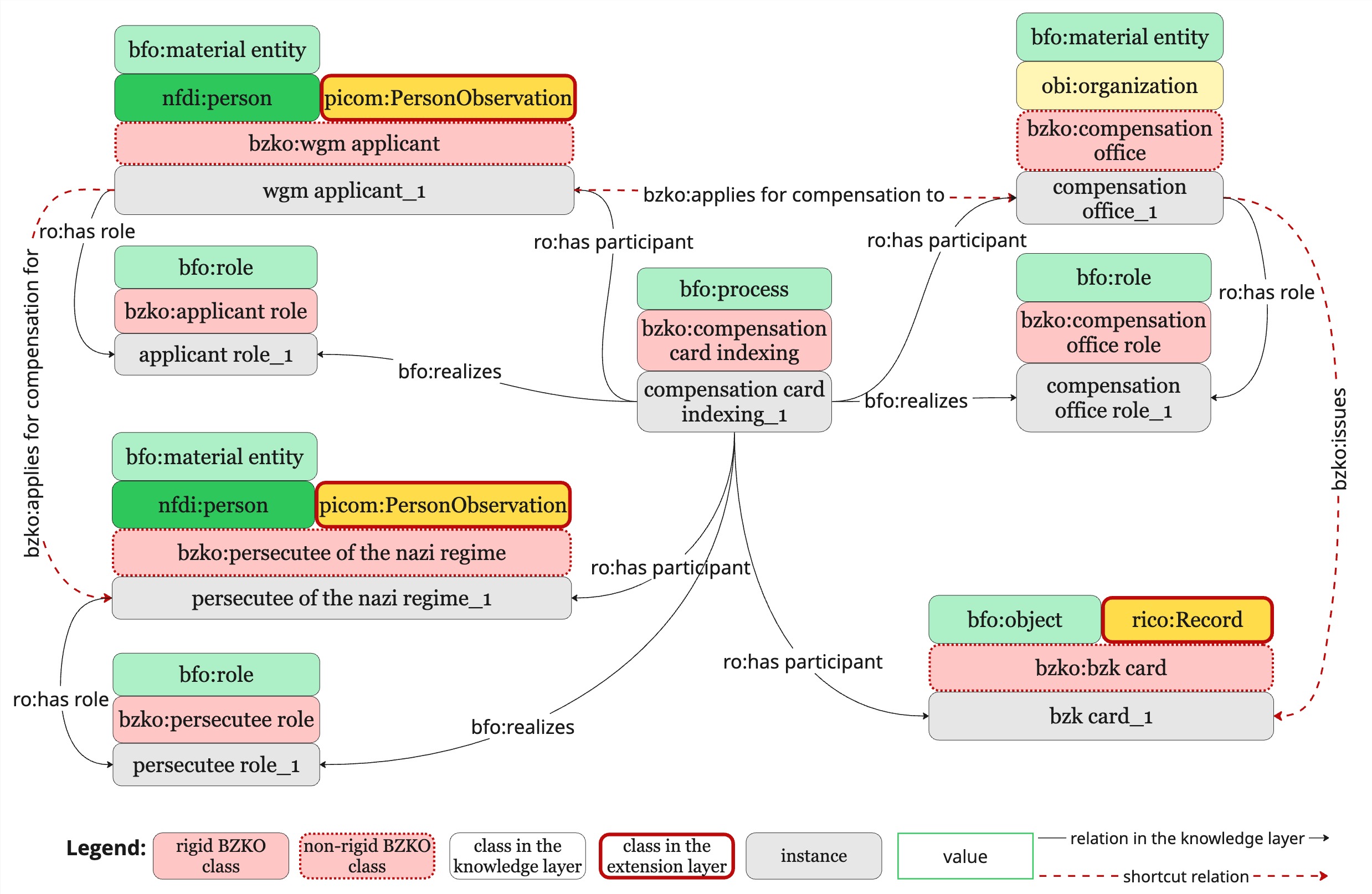}
  \caption{The process of compensation card indexing and its participants, along with all superclasses up to the closest BFO class in the taxonomy, the example instances, and the relations between them.}
    \label{fig:indexingCentered}
\end{figure}

To enhance the semantic representation of the BZK cards, the following relations between the classes are introduced: \textbf{bzko:applies for compensation for}, \textbf{bzko:applies for compensation to}, and \textbf{bzko:issues}. Figure~\ref{fig:indexingCentered} illustrates the domain and the range of these object properties, along with the imported and rigid and non-rigid classes of BZKO, with all superclasses up to the closest BFO class in the taxonomy. All the object and data properties in the BZKO knowledge layer are already defined in NDFIcore: they are either BFO, RO, or IAO relations, or NFDIcore relations.

The second group of non-rigid BZKO classes contains places. A place can be labeled in various forms, such as a birthplace, a hiding place, or a marketplace, but none of them are defined by the ontological status of the place itself (Cf. \cite{OntoClean}). A place, then, gains such extrinsic characteristics only within a context. The same holds for the third group of non-rigid BZKO classes, viz., dates. A date is modeled as a temporal instance; however, a date of birth is a temporal instance that depends on a birth process boundary and is therefore context-dependent. Consequently, \textit{bzko:birthplace} and \textit{bzko:deathplace}, and \textit{bzko:date of birth} and \textit{bzko:date of death} are non-rigid. 

The first group of relations, namely object properties of BZKO, is introduced to enhance the semantic representation of the BZK cards. As already illustrated in Figure~\ref{fig:indexingCentered}, \textbf{bzko:applies for compensation for},\textbf{ bzko:applies for compensation to}, and \textbf{bzko:issues} relate the classes to expose the domain knowledge beyond the content of the BZK cards. 
\begin{figure}
  \centering
  \includegraphics[width=\linewidth]{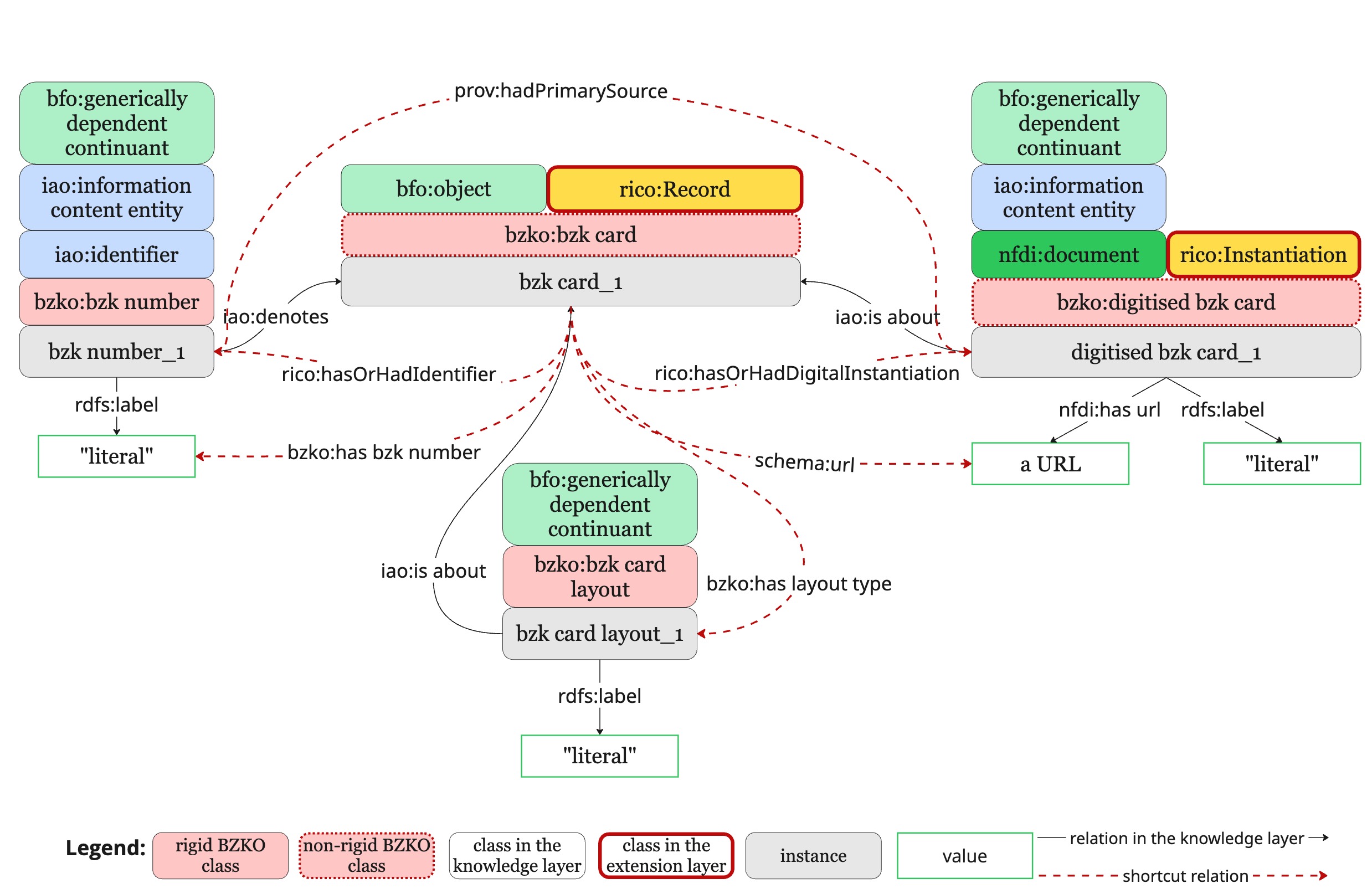}
  \caption{A diagram showing the relationships between a BZK index card, its digitized representation, its layout, and its number, as well as relationships in the extension layer, including the imported property \textbf{prov:hadPrimarySource} and a shortcut property \textbf{bzko:has layout type}. }
    \label{fig:digitalCard}
\end{figure}

The second group of relations, namely data properties of BZKO, is introduced for convenience, e.g., to enable direct access to literal values. These relations are redundant since the domain can already be represented using the existing entities. Figure~\ref{fig:digitalCard} illustrates the relations between a BZK index card, its digitized version, its layout, and its number. The relations, namely data properties, \textbf{bzko:has bzk number} and \textbf{bzko:has layout type} are added to facilitate semantic search, as they directly relate class instances to their data values. For the same purpose, other BZKO data properties are introduced to relate names and addresses to their corresponding literal strings. 

\subsection{The Extension Layer of BZKO}\label{extension layer}

The extension layer of BZKO includes (i) all classes and relations from the knowledge layer; namely, all rigid classes and primitive relations (Section~\ref{knowledgelayer}); (ii) the non-rigid classes and relations of BZKO (Section~\ref{nonrigid}); and (iii) classes and relations from the imported semantic resources PROV-O, RiC-O, PiCo, and Schema.org, widely used semantic standards in the digital humanities, to extend BZKO.

The classes \textit{rico:Record} and \textit{rico:Instantiation} are used to define \textit{bzko:bzk card} and \textit{bzko:digitized bzk card}, respectively, thereby assigning archival characteristics to the latter classes. BZKO also includes \textit{picom:PersonObservation}, a subclass of \textit{picom:Person}. Since \textit{picom:Person} refers to human beings, we interpret \textit{picom:PersonObservation} as referring to persons whose personal details are recorded in source documents. A more appropriate label would therefore be "Observed Person". Under this interpretation, a person mentioned on a BZK card instantiates both \textit{picom:PersonObservation} and \textit{nfdi:person}.

For the future integration of additional Wiedergutmachung documents, which also contain information about people, organizations, and addresses, into the Themenportal Wiedergutmachung, tracking the source of the data becomes crucial. In this respect, \textbf{prov:hadPrimarySource} is added to the ontology, and SPARQL \texttt{CONSTRUCT}  queries are defined to link the classes associated with data properties with the corresponding digitized BZK cards through this relation, hereby enabling instances and their associated property values to be traced back to their corresponding digitized BZK cards. Figure~\ref{fig:digitalCard} illustrates how these entities are used in the modeling, along with rigid and non-rigid classes and relations of BZKO. 

To express the relation between shortcut relations and their corresponding ontological patterns, we have defined a set of rules. The Semantic Web Rule Language (SWRL) rules were not used because they use transitive properties. As a result, not all reasoners, such as HermiT \cite{hermit}, fully support them. Moreover, SWRL rules may not allow the creation of new individuals, which is required when mapping shortcut relations to more complex ontological patterns. To ensure broad support across OWL reasoners, we defined rules in the form of SPARQL \texttt{CONSTRUCT} queries. These queries establish explicit mappings, allowing seamless translation in both directions; from shortcuts to full patterns and vice versa. By encoding these relationships as \texttt{CONSTRUCT} queries, we ensure that the mappings are formally defined, machine-readable, and consistently applicable across datasets based on the ontology. The same approach applies to complex mappings to external vocabularies, where \texttt{CONSTRUCT} queries are likewise used to define the rules governing the translation. As a simple example, consider \textbf{bzko:has bzk number} in Figure~\ref{fig:digitalCard}. Assuming all the prefixes are provided, which are available in the OWL file, a \texttt{CONSTRUCT}  query is designed to retrieve the BZK number directly from the card as follows.  

\begin{table}[ht]
\centering
{\ttfamily
\begin{tabular}{ll}
CONSTRUCT \{ &  \\
\hspace{3mm}    ?bzk\_card1 bzko:BZK\_0000003 ?label .\} & \#has bzk number \\
WHERE \{ & \\
\hspace{3mm}    ?bzk\_card1 a bzko:BZK\_0000002: . & \#bzk card \\
\hspace{3mm}    ?bzk\_number1 a bzko:BZK\_0000003 . & \#bzk number \\
\hspace{3mm}    ?bzk\_number1 iao:denotes ?bzk\_card1 . & \\
\hspace{3mm}    ?bzk\_number1 rdfs:label ?label . & \\
\}
\end{tabular}
}
\end{table}

\section{Discussion}

The BZK Ontology is being developed not merely as a graph representation of the OpenBZK, but as a semantic model that captures entities underlying the BZK cards. It aims at balancing interoperability with domain-specific expressiveness by being BFO-compliant and reusing prominent semantic resources in digital humanities, viz., RiC-O, PiCo, and PROV-O. As a result, it enables representation of archival data beyond what is explicitly stored in the database and also the maintenance of traceability to the contents of the cards. This traceability is implemented by linking all data values (agent names, dates, places, BZK numbers, addresses) to \textit{bzko:digitized bzk card} via \textbf{prov:hadPrimarySource}.

Although \textbf{prov:wasDerivedFrom}, which is the superproperty of \textbf{prov:hadPrimarySource}, has been subsumed by \textbf{ro:causally influenced by} in \cite{ro_bfo} (Cf. \cite{provo_bfo_mapping}), we do not adopt the latter relation in the knowledge layer of BZKO. On the one hand, causality is the central notion underlying \textbf{ro:causally related to} and its subproperties \cite{ROCausalDefn}, on the other hand, provenance concerns linking the data values to the digitized BZK cards. In BZKO, thus, provenance relations are not used in the sense of causal influences (Cf. \cite{ROCausalDoc}).

At present, \textit{bzko:compensation card indexing} models the creation of the card and identification of agents. A fuller model should distinguish among index card creation, application submission, and prosecution-related processes, and include relations that link each realizable entity to the process in which it is realized. A more fine-grained representation of these processes is left for future work.

BZK card layouts are modeled as named individuals in the ontology for two reasons. First, we cannot speak of an instance of a layout type: the design of the card, which is contingent, determines the layout type rather than instantiating it. Second, the archivists created a list of criteria to determine the card layout labels and provided a ground truth \cite{BZKopen} in which cards are associated with their corresponding layout. This suggests that cards are mapped to an item of a list.

In the ideal case, birth and/or death dates are modeled using \textbf{time:inXSDDate}, with values in the "YYYY-MM-DD" format. However, our review of the cards reveals that dates of birth and/or death may contain uncertain dates (e.g., "unbekannt" (\textit{Eng}. "unknown")), approximate dates (e.g., "ca. 1900"), incomplete dates (e.g., "Im Juli 1942"), or ambiguous dates (e.g., "1943 oder 1944", "14/26. März 1880", "75.70.1953", "1886–1865"). The ontology, then, must be extended to semantically represent all these cases. However, this is left as future work.

Some values recorded in the deathplace field do not denote a determinate place: they may refer to events (e.g., "in Deportation"), death-related annotations (e.g., "verstorben" (\textit{Eng}. "deceased")), or place names qualifed by uncertanty (e.g., "wahrscheinlich Auschwitz" (\textit{Eng}. "probably Auschwitz")), rather than plain assertions of location. Therefore, representing deathplaces as structured locations would not adequately capture the data. Consequently, these values are currently modeled as literal strings.

In the knowledge layer, we chose \textbf{ro:has participant} over \textbf{bfo:has participant} (Cf. \cite{ro_bfo}), as the domain of the former covers \textit{bfo:process boundary}, which is required to model birth and death as occurrences. Furthermore, \textit{bzko:compensation office name} is subsumed under \textit{iao:institutional identification}. However, its definition ("A textual entity intended to identify a particular institution") appears to suggest that it would more appropriately be classified under \textit{iao:identifier}, rather than \textit{iao:textual entity}. We adhere to the IAO design choices, although we find this modeling decision debatable. 

We intentionally avoid committing imported classes to specific BFO branches, since they do not provide a sufficiently specific BFO commitment. For instance, \textit{rico:Record}, the superclass of \textit{rico:RecordResource}, may refer to either an information content entity, as an instance of \textit{bfo:generically dependent continuant}, or a material entity, as an instance of \textit{bfo:independent continuant}. Similarly, \textit{picom:PersonObservation} refers to the persons who are observed on a \textit{picom:Source}, which is, again, can be an information content entity or a material entity. Thus, in order to avoid future implementation errors and inconsistencies, these classes are maintained as mapping-layer constructs rather than integrated into the BFO taxonomy.

The high-level patterns in the knowledge layer are not immediately clear when viewing the ontology. So, we introduced equivalent classes and shortcuts in the extension layer to compact complex expressions into simpler ones \cite{mungall_shortcut}. The BZKO relations, covering \textbf{bzko:issues}, \textbf{bzko:has layout type}, and \textbf{bzko:has alternative given name}, are introduced as shortcuts. Although in the knowledge layer complex multihop relation paths already express the semantics, they are not easy to understand for a non-ontology expert. Thus, BZKO relations are introduced to not only improve usability and readability by providing a straightforward, natural-language alignment but also to facilitate faster querying.

The extension layer also contains imported relations from RiC-O, PROV-O, and PiCo as shortcuts. These relations are preserved in the ontology when their semantics cannot be fully represented by the corresponding relations in the knowledge layer. For instance, \textbf{rico:hasOrHadInstantiation} extends the semantics of the model by characterizing a digitized BZK card as an archival instantiation rather than merely relating an instance of \textit{bfo:object} to an instance of \textit{iao:information content entity}. Similarly, \textbf{rico:hasOrHadIdentifier}, defined as a relation that "connects a Thing to one of its past or present Identifiers", whose definition reads "[c]onnects a [owl:]Thing to one of its past or present [rico:]Identifiers.", was not modeled as a subproperty of \textbf{iao:denoted by}, since the label of the RiC-O relation explicitly includes temporal aspects of identification, a separate mapping was provided in order to preserve its intended semantics. Additionally, to the best of our understanding ot the RiC-O documentation \cite{rico}, the \textit{hasOrHad-} naming convention used throughout RiC-O denotes a relation that may hold now or may have held at some point in the past, rather than committing to a specific formal theory of time. As such, these relations are compatible with the BFO framework, as they do not introduce any OWL DL temporal semantics. 

\section{Conclusion and Future Work }
This paper illustrates the development of BZKO, a BFO-compliant ontology for index cards in the Wiedergutmachung documents. The ontology distinguishes between a knowledge layer, which captures ontologically grounded entities and relations, and an extension layer, which introduces shortcuts and mappings for imported classes and relations from RiC-O, PROV-O, and PiCo to preserve archival semantics, support provenance tracking, and facilitate integration with existing digital humanities infrastructures. Thus, we aimed at balancing ontological rigor with archival practicality.

The development of the BZK Ontology is still ongoing, and three directions for future work remain: ontology validation, knowledge graph generation and data enrichment, as well as an extension of the domain coverage.

The ontology requires further validation through competency questions. To this end, meetings with archivists are being conducted to refine and complete the competency questions on validation and use cases. Once these have been established, the ontology can be systematically evaluated against the underlying database. The resulting competency questions will also provide a foundation for developing SWRL rules and SPARQL queries. 

Several research activities are currently underway within the \textit{Themenportal Wiedergutmachung} Project. For instance, address parsing, the correction of place names, modeling vague spatial references, and the normalization and linking of location information to GeoNames are currently under investigation. These developments will require corresponding extensions to the ontology, particularly regarding the representation of historical places and geographical entities. As the results of these studies become available, the ontology will be revised to accommodate the newly identified requirements.

Similarly, ongoing research on the normalization of temporal information will inform future revisions of the ontology. The current model represents dates as literals and does not yet capture the full range of uncertain, incomplete, approximate, and ambiguous temporal data found on the index cards. Future results in temporal data normalization will therefore guide the extension of the ontology's temporal representation.

Finally, as mentioned above, the BZK cards contain additional information that has not yet been modeled. Examples include familial relationships between applicants and persecutees, nationality, and companies, political parties, trade unions, religious communities, and similar organizations appearing as applicants and/or persecutees. Incorporating these entities and relations will require further extensions of both the ontology and the underlying database, as well as the formulation of new competency questions. Although these developments are considered long-term objectives, they remain part of the broader research agenda.	

\begin{acknowledgments}
This work is funded by the German Federal Ministry of Finance
(\textit{Bundesministerium der Finanzen}). We are indebted to Inger Louise Banse-Strobel, Kevin Dubout, and Dominik Sturm for the many valuable discussions on the archival domain, for their contributions to the development of the competency questions, and for the comments on the earlier version of this paper. We thank the anonymous reviewers for their valuable comments and constructive suggestions, which helped improve the quality and clarity of this paper.
\end{acknowledgments}

\section*{Declaration on Generative AI}
 During the preparation of this work, the authors used Grammarly in order to: Grammar and spelling check. After using this tool, the authors reviewed and edited the content as needed and take full responsibility for the publication’s content. 

\bibliography{references_bzko}

@String{Computing = "Computing" }

@String{Springer = "Springer-Verlag" }

@misc{BFM,
key = {BMF},
author = {{German Federal Ministry of Finance}},
year = 2026,
title = "Wiedergutmachung-Provisions relating to compensation for National Socialist injustice",
url = "https://www.bundesfinanzministerium.de/Content/EN/Standardartikel/Press_Room/Publications/Brochures/2018-08-15-entschaedigung-ns-unrecht-engl.pdf?__blob=publicationFile&v=3",
lastaccessed = "June 15, 2026",
}

@misc{ArchivPortal,
key =          {Archivportal-D},
year =         2024,
title =        "Themenportal Wiedergutmachung",
url =          "https://www.archivportal-d.de/themenportale/wiedergutmachung",
lastaccessed = "June 15, 2026",
}

@misc{BMFArchProject,
key =          {German Federal Ministry of Finance},
year =         2021,
title =        "Archivierungsprojekt Wiedergutmachung",
url =          "https://www.bundesfinanzministerium.de/Monatsberichte/2021/01/Inhalte/Kapitel-3-Analysen/3-7-archivierungsprojekt-wiedergutmachung.html",
note =         {{Monthly Report January 2021, in German}},
lastaccessed = "June 15, 2026",
}

@misc{BezirkDuesseldorf,
  key =          {Düsseldorf District Government},
  year =         2023,
  title =        {{Information Sheet on Compensation for Nazi Injustice (English)}},
  howpublished = {\url{https://www.brd.nrw.de/document/20230201_1_15_BZK_Merkblatt_englisch.pdf}},
  note =         {Available via the Compensation for Nazi Injustice information page: {\url{https://www.brd.nrw.de/Ueber-uns/Die-Bezirksregierung/Entschaedigung-fuer-Naziunrecht}}},
lastaccessed = "June 15, 2026",
}

@misc{BZKopen,
  key =          {BZKopen},
  author =       {Mahsa Vafaie},
  year =         2025,
  title =        "BZKopen Dataset",
  url =          "https://huggingface.co/datasets/MahsaVafaie/BZKopen",
  note =         {{Hugging Face dataset repository}},
  lastaccessed = "June 15, 2026",
}

@book{Arp2015BFO,
author    = {Robert Arp and Barry Smith and Andrew D. Spear},
title     = {Building Ontologies with Basic Formal Ontology},
publisher = {MIT Press},
address   = {Cambridge, MA},
year      = {2015}
}

@misc{ISOBFO,
author       = {{International Organization for Standardization}},
title        = {{ISO/IEC 21838-2:2020 Information Technology --- Top-Level Ontologies (TLO) --- Part 2: Basic Formal Ontology (BFO)}},
year         = {2020},
url          = {https://www.iso.org/standard/74572.html},
lastaccessed = "June 15, 2026",
}

@misc{NFDI,
year =         2024,
author  = {{National Research Data Infrastructure}},
url =          "https://www.nfdi.de",
lastaccessed = "June 15, 2026",
}

@misc{NFDICoreGithub,
key =          {NFDIcore Ontology GitHub Repository},
year =         2025,
url =          "https://github.com/ISE-FIZKarlsruhe/nfdicore",
lastaccessed = "June 15, 2026",
}

@misc{BFOGithub,
key =          {Basic Formal Ontology GitHub Repository},
year =         2025,
url =          "https://github.com/bfo-ontology",
lastaccessed = "June 15, 2026",
}

@inproceedings{vafaie2025end,
  title={End-to-end Information Extraction from Archival Records with Multimodal Large Language Models},
  author={Vafaie, Mahsa and Hertling, Sven and Banse-Strobel, Inger and Dubout, Kevin and Sack, Harald},
  booktitle={Proceedings of the 34th ACM International Conference on Information and Knowledge Management},
  pages={6075--6083},
  year={2025},
    doi = {10.1145/3746252.3761503},
}

@misc{OBO,
key =          {Open Biological and Biomedical Ontology Foundry Principles},
year =         2024,
url =          "https://obofoundry.org/principles/fp-000-summary.html",
lastaccessed = "June 15, 2026",
}

@article{OBOFoundry2021,
    author = {Jackson, Rebecca and Matentzoglu, Nicolas and Overton, James A and Vita, Randi and Balhoff, James P and Buttigieg, Pier Luigi and Carbon, Seth and Courtot, Melanie and Diehl, Alexander D and Dooley, Damion M and Duncan, William D and Harris, Nomi L and Haendel, Melissa A and Lewis, Suzanna E and Natale, Darren A and Osumi-Sutherland, David and Ruttenberg, Alan and Schriml, Lynn M and Smith, Barry and Stoeckert Jr., Christian J and Vasilevsky, Nicole A and Walls, Ramona L and Zheng, Jie and Mungall, Christopher J and Peters, Bjoern},
    title = {OBO Foundry in 2021: operationalizing open data principles to evaluate ontologies},
    journal = {Database},
    volume = {2021},
    pages = {baab069},
    year = {2021},
    month = {09},
    issn = {1758-0463},
    doi = {10.1093/database/baab069},
}

@article{mungall_shortcut,
  title={Taking shortcuts with OWL using safe macros},
  author={Mungall, Christopher and Ruttenberg, Alan and Osumi-Sutherland, David},
  journal={Nature Precedings},
  pages={1--1},
  year={2010},
  publisher={Nature Publishing Group UK London},
doi = {https://doi.org/10.1038/npre.2010.5292.1},
}

@article{ODK_paper,
    author = {Matentzoglu, Nicolas and Goutte-Gattat, Damien and Tan, Shawn Zheng Kai and Balhoff, James P and Carbon, Seth and Caron, Anita R and Duncan, William D and Flack, Joe E and Haendel, Melissa and Harris, Nomi L and Hogan, William R and Hoyt, Charles Tapley and Jackson, Rebecca C and Kim, HyeongSik and Kir, Huseyin and Larralde, Martin and McMurry, Julie A and Overton, James A and Peters, Bjoern and Pilgrim, Clare and Stefancsik, Ray and Robb, Sofia MC and Toro, Sabrina and Vasilevsky, Nicole A and Walls, Ramona and Mungall, Christopher J and Osumi-Sutherland, David},
    title = {Ontology Development Kit: a toolkit for building, maintaining and standardizing biomedical ontologies},
    journal = {Database},
    volume = {2022},
    pages = {baac087},
    year = {2022},
    month = {01},
    issn = {1758-0463},
    doi = {10.1093/database/baac087},
}

@misc{ODKGitHub,
key =          {ODK},
author = {{INCATools}},
year =         2024,
title =        "Ontology Development Kit",
url =          "https://incatools.github.io/ontology-development-kit/",
lastaccessed = "June 15, 2026",
}

@inproceedings{CourtDoc_paper,
author = {Vafaie, Mahsa and Bruns, Oleksandra and Pilz, Nastasja and Waitelonis, J\"{o}rg and Sack, Harald},
title = {CourtDocs Ontology: Towards a Data Model for Representation of Historical Court Proceedings},
year = {2023},
isbn = {9798400701412},
publisher = {Association for Computing Machinery},
address = {New York, NY, USA},
doi = {10.1145/3587259.3627562},
booktitle = {Proceedings of the 12th Knowledge Capture Conference 2023},
pages = {175–179},
numpages = {5},
location = {Pensacola, FL, USA},
series = {K-CAP '23}
}

@article{OntoClean,
  title={An overview of OntoClean},
  author={Guarino, Nicola and Welty, Christopher A},
  journal={Handbook on ontologies},
  pages={201--220},
  year={2009},
  publisher={Springer}
}

@techreport{rico,
author      = {{Expert Group on Archival Description (EGAD)}},
title       = {{Records in Contexts Ontology (RiC-O) Version 1.1}},
institution = {International Council on Archives},
year        = {2023},
url         = {https://www.ica.org/standards/RiC/RiC-O_1-1.html},
lastaccessed = "June 15, 2026",
}

@misc{pico,
  author =       {Ivo Zandhuis and Jeroen Balkenende and Pieter Woltjer and Bob Coret and Mark Lindeman},
  title =        {{Persons in Context (PiCo)}},
  institution =  {Centre for Family History},
  year =         {2026},
  version =      {1.1.0},
  url =          {https://personsincontext.org/model/},
  lastaccessed = {June 15, 2026},
}

@misc{ro_bfo,
author = {{Relations Ontology Project}},
title  = {{Relations in RO and BFO}},
year   = {2026},
url    = {https://github.com/oborel/obo-relations/blob/8154e83fd3e1b5203929dda24adbd8983ac00715/docs/ro-and-bfo.md},
lastaccessed = "June 15, 2026",
}

@misc{provo_bfo_mapping,
  author       = {Prudhomme, Tim and
                  De Colle, Giacomo and
                  Liebers, Austin and
                  Sculley, Alec and
                  Xie, Peihong and
                  Cohen, Sydney and
                  Beverley, John},
  title        = {PROV-to-BFO: v2025-01-19},
year   = {2025},
note =         {{[Data set] Zenodo]}},
  doi          = {10.5281/zenodo.14692262},
lastaccessed = "June 15, 2026",
}

@misc{ROCausalDefn,
  key =          {Relations Ontology},
  title =        {RO: causally related to (RO\_0002410)},
url =          "http://purl.obolibrary.org/obo/RO_0002410",
year =         2024,
lastaccessed = "June 15, 2026",
}

@misc{ROCausalDoc,
  key =          {Relations Ontology},
  year =         2025,
  title =        {Causal Relations in the OBO Relation Ontology},
  url =          {https://github.com/oborel/obo-relations/blob/master/docs/causal-relations.md},
  lastaccessed = {June 15, 2026},
}

@article{hermit,
  title={HermiT: an OWL 2 reasoner},
  author={Glimm, Birte and Horrocks, Ian and Motik, Boris and Stoilos, Giorgos and Wang, Zhe},
  journal={Journal of automated reasoning},
  volume={53},
  number={3},
  pages={245--269},
  year={2014},
  publisher={Springer},
  doi          = {10.1007/s10817-014-9305-1},

}

@inproceedings{vafaie2021modelling,
  title={Modelling archival hierarchies in practice: Key aspects and lessons learned},
  author={Vafaie, Mahsa and Oleksandra, Bruns and Pilz, Nastasja and Dess{\'\i}, Danilo and Sack, Harald and others},
  booktitle={CEUR workshop proceedings},
  volume={2981},
  year={2021},
  organization={CEUR-WS}
}

\end{document}